\documentstyle[twoside,fleqn,espcrc2,epsf]{article}
\input{epsf}

\title{Vortex Pinning and Dynamics in Layered Superconductors with Periodic
Pinning Arrays} 
\author{Charles Reichhardt $^{\rm a}$, Cynthia J. Olson 
\address{Department of Physics, University of California, Davis, California
95616}
and
Niels Gr{\o}nbech-Jensen
\address{Department of Applied Science, University of California, Davis, 
California 95616}
\address{NERSC, Lawrence Berkeley National Laboratory, Berkeley, 
California 94720}
}

\begin{document}

\begin{abstract}
We examine vortex dynamics and pinning in layered superconductors
using three-dimensional molecular dynamics simulations of magnetically
interacting pancake vortices. 
Our model treats the
magnetic interactions of the pancakes exactly, with long-range logarithmic
interactions both within and between planes.  
At the matching field the vortices are aligned with the pinning array. 
As a function of tilt angle for the pinning arrays a series of commensuration
effects occur, seen as peaks in the critical current, due to
pancakes finding a favorable alignment.
\vspace{1pc}
\end{abstract}

\maketitle

In superconductors with periodic pinning arrays interesting commensurability
effects occur when
the periodicity of the vortex lattice matches the 
periodicity of the pinning lattice. 
Experiments \cite{Baert,Schuller} 
and simulations \cite{Reichhardt} 
so far have been done with
thin film superconductors where the vortex lattice and pinning 
can be considered two-dimensional. The case of vortex lattices interacting
with a periodic pinning array in a layered 3D superconductor has not been
studied. Such a system would correspond to an anisotropic superconductor
such as BSCCO  
with a periodic arrangement of columnar defects. In this system   
the $z$-direction
becomes important as the applied field or the pinning array is tilted. 
The dynamical effects of
vortices moving in periodic pinning arrays 
in such a system have not been examined,
in particular
how the  vortex lattice structure of the moving state differs from that
of the pinned state. 
To study vortex pinning and 
dynamics in layered superconductors,
we have developed a simulation containing
the correct magnetic interactions between pancakes \cite{clem}.
This interaction is long range both in and between planes, and is
treated using a rapidly converging summation method \cite{ngj}. 

The overdamped equation of motion, for $T=0$, for vortex $i$ is given by
$ {\bf f}_{i} = \sum_{j=1}^{N_{v}}\nabla {\bf U}(\rho_{i,j},z_{i,j})
+ {\bf f}_{i}^{vp} + {\bf f}_{d}= {\bf v}_{i}$,
where $N_v$ is the number of vortices and $\rho$ and $z$ are the distance
between pancakes in cylindrical coordinates.
The magnetic energy between pancakes is 
\begin{eqnarray}
{\bf U}(\rho_{i,j},0)=2d\epsilon_{0} 
\left((1-\frac{d}{2\lambda})\ln{\frac{R}{\rho}}
+\frac{d}{2\lambda} 
E_{1}(\rho)
\right) 
\nonumber
\end{eqnarray}
\begin{eqnarray}
{\bf U}(\rho_{i,j},z)=-\frac{d^{2}\epsilon_{0}}{\lambda}
\left(\exp(-z/\lambda)\ln\frac{R}{\rho}- 
E_{1}(R)
\right) \nonumber
\end{eqnarray}
where
$R = \sqrt{z^2 + \rho^2}$, 
$E_{1}(x) = 
\int^{\infty}_{\rho}\exp(-x/\lambda)/\rho^{\prime}d\rho^{\prime}$ and
$\epsilon_{0} = \Phi_{0}^{2}/(4\pi\xi)^{2}$.
The pinning is placed in a square array of parabolic traps with 
a radius $r_{p}$ much smaller than the distance between pins. 
The location of the pinning sites is the same in every layer corresponding
to correlated defects. 
A driving force $f_{d}$ is slowly increased and the vortex velocities are 
measured. 
Here we consider the 
first matching field
case where the
number of vortices $N_{v}$ equals the number of pinning sites $N_{p}$. 
We conduct a series of 
simulations in which the pinning sites are tilted 
at an increasing angle
with respect to the 
$z$-axis. We will only consider driving 
that produces vortex motion
transverse to the direction of the tilt angle. 
We examine systems with 8 layers 
containing 64 vortices and pins in each layer. 
Work for larger systems, varied 
fields and coupling strength will be presented elsewhere \cite{toappear}.

In Fig. 1(a) we present 
the critical depinning force $f_{dp}^{c}$ 
as a function of tilt angle $ \theta$. 
Here $f_{dp}^{c}$ peaks 
at $\theta = 0^{\circ}$ when the 
pancakes are aligned with pins on all layers. 
As $ \theta$ is increased
$f_{dp}^{c}$ drops. 
For small tilt angles $ \theta < 5^{\circ}$ 
the vortex lines tilt with the pins. For larger angles the vortex lines
realign in the $z$ direction.  
The depinning force $f_{dp}^{c}$ 
will then remain low 
as only one pancake in the straight vortex line will 
be sitting at a pinning site. At 
$\theta = 45^{\circ}$ $f_{dp}^{c}$ shows a peak of the same magnitude 
as the peak at 
$\theta = 0$. At this tilt angle, and also for any angle
satisfying 
$ \theta = \tan^{-1}(n)$ where $n$ is an 
integer, the pinning sites are again
aligned in the $z$-direction so that a vortex line can be formed that
is also aligned in the $z$-direction with 
all the pancakes in a single vortex being able to 
sit in a pinning site.  
There are also peaks in $f_{dp}^{c}$ at 
$\theta = 26.6^{\circ}$ and $56.3^{\circ}$.
At these angles the pancakes again sit on all the pinning sites.
The individual vortex lines now consists of half the number of pancakes
as at $\theta = 0.0^{\circ}$; however, there are now twice as many vortex
lines with the pancakes from an individual vortex line being coupled in
every other layer. 
The view from the
$z$-direction as shown in Fig.~1 for these angles  
indicates that the vortex lattice is now rectangular with
twice as many vortex lines as at the other angles. 
At $\theta = 36.9^{\circ}$ a smaller peak is observed. The vortex structure
at this angle will be presented elsewhere \cite{toappear}. 
 
In (b) and (c) we show the vortex structures for the pinned phase and
moving phase for $\theta = 1.5^{\circ}$ as seen from the $z$-direction. 
In (b) the vortices can be seen to stay aligned with the pins. In (c)
for $f_{d} > f_{dp}^{c}$ the vortices realign with the z-direction.  
Such a transition from a tilted to straight vortex lattice as a function
of drive may be visible with neutron scattering experiments.

We acknowledge helpful discussions with L. N. Bulaevskii, A. Kolton,
R.T. Scalettar, and  G. T. Zim{\' a}nyi. 
This work was supported by CLC and CULAR (LANL/UC) and by the Director,
Office of Adv. Scientific Comp. Res., Div. of Math., Information and
Comp. Sciences, U.S.~DoE contract DE-AC03-76SF00098.

\begin{figure}
\centerline{
\epsfxsize=7cm 
\epsfbox{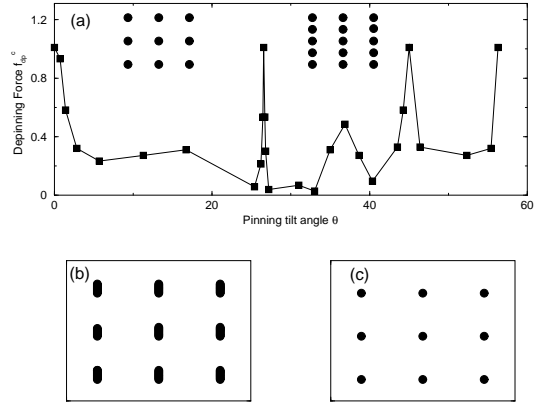}}
\caption{The critical depinning force $f_{dp}^{c}$ versus the tilt angle
$\theta$ of the pinning sites. The vortex arrangements as seen from the
$z$-direction are outlined for different tilt angles $\theta = 0$ left and
$\theta = 26.6$ right.
(b) shows the pinned vortex arrangement for $\theta = 1.5^{\circ}$ where
the vortices stay aligned with the pins. (c) shows the moving vortex 
state for $\theta = 1.5^{\circ}$ 
where the vortices have realigned with the $z$ direction.}
\label{fig:fig1}
\end{figure}

\end{document}